\documentclass[11pt]{article}
\usepackage{amssymb}
\usepackage{bbm}
\usepackage{tabularx}
\usepackage{array}
\usepackage{makecell}
\usepackage{tabularx}
\usepackage{pgfplots}
\usepgfplotslibrary{groupplots}
\usepackage{pgfplotstable}
\pgfplotsset{compat=1.18}
\usepackage[final]{acl}
\usepackage{times}
\usepackage{enumitem}
\usepackage{latexsym}
\usepackage{amsmath}
\usepackage[T1]{fontenc}
\usepackage[utf8]{inputenc}
\usepackage{microtype}
\usepackage{inconsolata}
\usepackage{graphicx}

\title{Evaluating LLMs as Human Surrogates in Controlled Experiments}

\author{Adnan Hoq \\
  University of Notre Dame \\
  Notre Dame, Indiana, USA \\
  \texttt{ahoq@nd.edu} \\\And
  Tim Weninger \\
  University of Notre Dame \\
  Notre Dame, Indiana, USA \\
  \texttt{tweninger@nd.edu} \\}

\begin{document}
\maketitle

\begin{abstract}

Large language models (LLMs) are increasingly used to simulate human responses in behavioral research, yet it remains unclear when LLM-generated data support the same experimental inferences as human data. We evaluate this by directly comparing off-the-shelf LLM-generated responses with human responses from a canonical survey experiment on accuracy perception. Each human observation is converted into a structured prompt, and models generate a single 0--10 outcome variable without task-specific training; identical statistical analyses are applied to human and synthetic responses. We find that LLMs reproduce several directional effects observed in humans, but effect magnitudes and moderation patterns vary across models. Off-the-shelf LLMs therefore capture aggregate belief-updating patterns under controlled conditions but do not consistently match human-scale effects, clarifying when LLM-generated data can function as behavioral surrogates.

\end{abstract}

\section{Introduction}

Large language models (LLMs) are increasingly used not only as generative systems, but as tools for simulating human behavior~\cite{messeri2024artificial,aher2023using,horton2023large,argyle2023out,binz2023using,kozlowski2025simulating,park2023generative,horton2023large}. They are treated as ``silicon samples'' that can stand in for people by producing plausible responses in surveys, experiments, and interactive settings~\cite{messeri2024artificial,kozlowski2025simulating,argyle2023out,cui2024can,li2023camel,hullman2025human}. In practice, a short textual persona, including demographics, political identity, context, or experimental condition, is given to a model, which then generates that person’s likely response~\cite{hullman2025human,cui2025large,argyle2023out,aher2023using,park2023generative}. Can realistic human responses really be produced from such minimal descriptions? If so, human-like responses could be generated on demand, with implications for how opinions are measured, policies are evaluated, and social behavior is studied at scale~\cite{gilardi2023chatgpt,tornberg2023chatgpt,hullman2025human}.

This idea is ambitious. Behavioral science has long grappled with high data-collection costs, limited statistical power, replication crises, and challenges in cumulative theory-building~\cite{open2015estimating,camerer2018evaluating,smaldino2016natural,muthukrishna2019problem,munafo2017manifesto}. LLM-based surrogates promise relief from these constraints. They enable rapid exploration of design variations, counterfactual conditions, and hard-to-reach populations~\cite{li2023camel,li2023counterfactual,argyle2023out,park2023generative}. They also make it possible to simulate historical contexts and elite or rare subgroups~\cite{hua2023war,gao2024large,li2023camel}. They could support large-scale experimentation without respondent fatigue~\cite{gilardi2023chatgpt,aher2023using}. In short, LLMs are being used as flexible ``what-if'' engines for behavioral inquiry.

\subsection*{Does LLM-behavior match humans-behavior?}

At the same time, the use of LLMs as behavioral surrogates has prompted substantial methodological and philosophical debate~\cite{wang2025large}. Empirical studies show that LLM-generated responses often approximate patterns observed in human data, consistent with their training on large corpora of human-produced text~\cite{aher2023using,argyle2023out,abdurahman2024perils,bisbee2024synthetic}. Recent large-scale replication efforts report that GPT-4 reproduces between 73\% and 81\% of published main effects across dozens of psychological experiments, and 46\% to 63\% of interaction effects~\cite{cui2025large,cui2024can}. Linguistic and semantic studies similarly document strong correlations between LLM and human judgments~\cite{chiang2023can,elangovan2024beyond,bavaresco2025llms,dominguez2024questioning}.

Replication results are uneven. LLMs often produce larger effect sizes than humans, yield significant results where original studies reported null findings, and show sensitivity to socially charged domains such as race or gender~\cite{cui2025large,cui2024can}. These discrepancies raise concerns about effect-size inflation, systematic bias, and overconfidence. Existing evaluations often either (i) establish directional similarity under related settings or (ii) apply calibration procedures that adjust LLM outputs using gold-standard human data~\cite{messeri2024artificial}.

A central question remains: when does LLM-generated data support the same hypothesis tests as human data? Directional replication and correlational convergence are often observed, but preservation of experimental inferences under identical statistical models is uncertain. In other words, can off-the-shelf LLMs reproduce the empirical conclusions of a controlled experiment, rather than merely generate responses that look plausible in aggregate?

\subsection*{Our approach: structural agreement under identical analysis}

We test whether off-the-shelf closed- and open-source LLMs reproduce the same hypothesis-level conclusions as a human behavioral experiment under identical statistical analysis. We use a canonical instance of LLM-based behavioral simulation: \textit{belief judgments in a controlled information experiment}. Human participants rated the perceived accuracy of political news headlines under experimentally manipulated conditions~\cite{pfander2025spotting} (headline only versus headline plus AI credibility feedback~\cite{li2024impact}). Each human observation is converted into a structured prompt containing a persona description, the assigned condition, and the headline text. The model produces a single 0--10 perceived-accuracy rating without task-specific training, calibration, or prior response history. We then apply the same statistical analyses to human and synthetic data and compare the resulting hypothesis tests.

Our target is agreement in experimental inference. A valid surrogate should preserve the same treatment effects, ideological ordering, and stimulus-level variation observed in humans under identical analysis. Failures should appear as discrepancies in effect direction, effect magnitude, or moderation patterns.

\subsection*{Hypotheses}

We evaluate three structural hypotheses that characterize belief judgments in controlled information experiments:

\textbf{H1 (Political Alignment Effect).} Perceived accuracy varies systematically with ideological alignment between participant affiliation and headline content.

\textbf{H2 (Exposure Effect).} Credibility feedback produces a statistically significant shift in perceived accuracy relative to control.

\textbf{H3 (Headline-Level Heterogeneity).} Belief updating varies across headlines, indicating structured stimulus-level sensitivity rather than uniform shifts.

These hypotheses capture three orthogonal properties of the task: ideological ordering, causal treatment response, and stimulus-level variation. Valid behavioral surrogates must reproduce all three simultaneously under identical analysis; directional replication alone is insufficient.

\subsection*{Findings in Brief} 

Off-the-shelf LLMs reproduce several directional effects observed in humans. Ideological alignment patterns are often preserved (H1: partially confirmed), and credibility exposure produces statistically significant shifts in perceived accuracy across models (H2: confirmed). Effect magnitudes, however, vary substantially across systems: some models approximate human-scale treatment effects, while others exhibit exaggerated responsiveness. Headline-level heterogeneity is partially reproduced, with fidelity differing across model families (H3: partially confirmed).

LLMs therefore replicate aggregate behavioral regularities under controlled conditions, but quantitative alignment is model-dependent. Simply put, \textit{LLM-generated data are \textbf{not} interchangeable with human samples}; structural replication must be established hypothesis by hypothesis. By grounding evaluation in identical statistical specifications and effect-size comparisons, this framework identifies when off-the-shelf LLMs function as behavioral surrogates in social and behavioral research.

\section{Methods}

We compare human judgments and LLM-generated responses within the same experimental framework. Human data come from a repeated-measures experiment on news accuracy judgments. Participants rated the perceived accuracy of political news headlines under two conditions. In the control condition, headlines were presented without credibility signals. In the treatment condition, headlines were accompanied by an AI-generated credibility label indicating assessed accuracy.

\subsection{Data and Recruitment}
Participants were recruited via Prolific and screened for English fluency and regular news consumption. The analyses reported here draw on the Control (n = 278) and the Treatment (n = 244) groups, yielding a combined analytic sample of 522 participants. Assignment to conditions was conducted using block randomization to ensure balanced demographic representation across political affiliation, gender, race, age, education, and geographic residence (urban, suburban, rural). 

Details of human data collection procedures are provided in Appendix~\ref{app:human_data}.

\subsection{Human Judgment Data}
The human data used in this study come from an experiment examining news accuracy judgments. Participants were asked to rate the perceived accuracy of political news headlines on a fixed 0–10 numerical scale.

The design was between-subjects: each participant was randomly assigned to a single condition. In the Control condition, headlines were presented without any accompanying credibility information. In the Feedback condition, headlines were shown with an AI-generated credibility label indicating the system’s assessment of accuracy. Participants evaluated the same set of headlines within their assigned condition but did not view headlines across multiple conditions.
Each headline was displayed with social engagement metrics (likes, shares, and comments). These engagement counts were randomly generated and varied across participants.

After completing the task, participants filled out a post-experiment survey. The dataset includes response data (accuracy ratings), interaction/log data, and post-experiment survey data.

\paragraph{Treatment Condition: Credibility Labels.}
The feedback condition included discrete credibility labels with an ordered structure ranging from inaccurate to accurate (e.g., \textit{Inaccurate}, \textit{Unverified}, \textit{Somewhat Accurate}, \textit{Accurate}). Labels were intentionally categorical rather than probabilistic to approximate simplified credibility cues commonly presented on digital platforms and AI systems, and to enable tests of ordered belief updating rather than binary correction.

\paragraph{Political Affiliation.}
Participants self-reported political affiliation, used as a grouping variable in analysis. Affiliation is treated as a structural moderator of belief updating rather than a prediction target. Analyses examine whether ideological differences are preserved, amplified, or attenuated under AI feedback in both human and model-generated responses.

\subsection{LLM Surrogate Generation}

LLM responses are generated at the same observational granularity as the human experiment. Each observation, defined by a participant persona, headline, and experimental condition, is converted into a single model prompt. Control observations are presented without credibility feedback, and feedback observations include the same credibility label shown to participants. The model outputs one perceived-accuracy rating for each observation.

\paragraph{Persona Encoding.}
Each prompt includes a textual description of the participant constructed from available metadata. The core descriptor includes political affiliation. Where available, additional demographic attributes (\textit{e.g.}, age bracket, gender, education level) are included. No prior response history or individualized behavioral memory is provided. Ratings are therefore generated from static persona descriptors, assigned condition, and headline content.

\paragraph{Stimulus and Information Control.}
Headline text is provided verbatim to the model. Partisan-leaning annotations are excluded from prompts, so models receive the same information as participants. Leaning labels are used only for post-hoc analysis and hypothesis testing.

\paragraph{Prompt Structure and Decoding.}
Prompts follow a standardized template specifying the participant persona, headline text, and, when applicable, the AI credibility label. Models are instructed to output a single integer between 0 and 10 representing perceived headline accuracy. No additional explanation is permitted. Decoding is deterministic (temperature = 0) to eliminate sampling variability and ensure replicability.

\paragraph{Models.}
We evaluate both closed-source and open-source, off-the-shelf LLMs without task-specific fine-tuning. All models are queried using identical prompt templates and output constraints to ensure comparability. All models were run with temperature set to $0$ to ensure deterministic outputs. 

We evaluated three large language models as behavioral surrogates: \textbf{Llama~3.2:3B} ($\sim$3B parameters; Meta), \textbf{Gemma~2:9B} ($\sim$9B parameters; Google), and \textbf{GPT-5.2} (OpenAI, proprietary parameter count undisclosed). Llama and Gemma were deployed locally via Ollama, while GPT-5.2 was accessed through the OpenAI Responses API.

\subsection{Evaluation Strategy}

Model-generated ratings are analyzed using the same statistical specifications applied to the human data. All hypothesis tests are recomputed on synthetic responses using identical mixed-effects models and ANOVA designs, with the same moderators, contrasts, and effect-size measures.

Evaluation targets three structural properties corresponding to the study hypotheses:

\begin{enumerate}
    \item \textbf{Ideological alignment structure (H1).} Mixed-effects models assess whether perceived accuracy varies with headline alignment relative to participant affiliation and whether models reproduce the human ingroup–outgroup ordering and magnitude.

    \item \textbf{Credibility exposure effects (H2).} Two-way ANOVA evaluates whether AI feedback shifts perceived accuracy and whether exposure $\times$ affiliation moderation patterns match human results, comparing partial $\eta^2$ magnitudes across agents.

    \item \textbf{Headline-level belief updating (H3).} A headline-aware updating operator measures feedback-induced rating changes across headlines, affiliations, and labels. Structural fidelity is assessed via correlation with human $\Delta$ vectors, and magnitude fidelity via RMSE.
\end{enumerate}

This evaluation tests whether LLM-generated responses preserve the same experimental inferences as human data. Structural replication requires recovery of effect direction, interaction structure, and relative heterogeneity across stimuli, not merely similarity in aggregate rating distributions.

\section{Results}

We compare LLM-generated and human responses under identical statistical analyses to assess whether models reproduce the same experimental patterns. Results are organized by the three study hypotheses: ideological alignment effects (H1), credibility exposure effects (H2), and headline-level belief updating (H3).

\subsection*{Finding 1: Political Alignment Shapes Credibility Judgments and Differentiates Model Fidelity (H1)}

We examine whether perceived accuracy varies with the ideological alignment between a participant’s political identity and a headline’s leaning. Alignment is coded as \{\textit{Neutral}, \textit{Ingroup}, \textit{Outgroup}\} relative to the participant’s affiliation. Because ratings are repeated across both participants and headlines, mixed-effects models include crossed random intercepts for participant and headline.

\begin{table}[t]
\centering
\small
\renewcommand{\arraystretch}{1.1}
\begin{tabularx}{\columnwidth}{
    >{\centering\arraybackslash}X
    >{\centering\arraybackslash}X
    >{\centering\arraybackslash}X
    >{\centering\arraybackslash}X
    >{\centering\arraybackslash}X
    >{\centering\arraybackslash}X
}
\hline
\thead{Model}
& \thead{$\beta_I$}
& \thead{$\beta_O$}
& \thead{$\Delta_{IO}$}
& \thead{Ingrp\\Effect}
& \thead{Gap\\Ratio} \\
\hline
Human   & -0.701 & -1.200 & 0.499 & Yes & -- \\
GPT5.2 & -0.969 & -1.400 & 0.431 & Yes & 0.86 \\
Gemma   & -0.626 & -0.807 & 0.181 & Yes & 0.36 \\
Llama   & 0.003  & 0.005  & $\approx 0$ & No  & 0.00 \\
\hline
\end{tabularx}
\caption{Political alignment effects across humans and LLM surrogates. $\beta_I$ and $\beta_O$ are mixed-effects coefficients relative to Neutral headlines. $\Delta_{IO}=\beta_I-\beta_O$. Gap Ratio is relative to the human $\Delta_{IO}$. Ingrp Effect indicates whether Ingroup differs significantly from Neutral.}
\label{tab:h1_alignment}
\end{table}

\paragraph{Human Results.}
Humans exhibit a strong centrist preference: Neutral headlines receive the highest ratings, followed by Ingroup and then Outgroup content. The ingroup--outgroup gap is $\Delta_{IO}=0.499$ (Table~\ref{tab:h1_alignment}), indicating substantial ideological differentiation. Moderates rate headlines slightly higher (\textit{i.e.}, as more accurate) overall than Liberals and Conservatives.

\paragraph{Model Fidelity to Alignment Structure.}
Table~\ref{tab:h1_alignment} summarizes alignment effects across humans and models. Structural fidelity differs sharply:

\begin{enumerate}[leftmargin=*, labelwidth=2.2cm, labelsep=0.5em, align=left,
                  itemsep=0pt, parsep=0pt, topsep=2pt]
\item[\textbf{GPT 5.2}] Reproduces the human alignment pattern, with an ingroup--outgroup gap close to the human benchmark ($0.86\times$ human; Table~\ref{tab:h1_alignment}).

\item[\textbf{Gemma}] Preserves the directional ordering but compresses ideological differentiation, yielding a smaller gap ($0.36\times$ human).

\item[\textbf{Llama}] Shows no alignment sensitivity: neither Ingroup nor Outgroup differs from Neutral, and $\Delta_{IO}\approx0$.
\end{enumerate}

\paragraph{Summary.}
H1 is strongly supported in human data: center-aligned headlines receive the highest ratings, outgroup-aligned headlines the lowest, with a robust ingroup–outgroup gap. Among LLM surrogates, GPT 5.2 most closely matches both the direction and magnitude of human ideological differentiation, Gemma preserves ordering with attenuation, and Llama shows no alignment sensitivity. These results reveal pronounced heterogeneity in structural fidelity across models.

\begin{figure}[t]
\centering
\pgfplotstableread[col sep=comma]{
agent,eta2,star
Llama,0.167816,***
Gemma,0.000616,**
GPT 5.2,0.001614,***
Human,0.003135,***
}\hTwoMain

\pgfplotstableread[col sep=comma]{
agent,eta2,star
Llama,0.011000,***
Gemma,0.008000,***
GPT 5.2,0.000010,
Human,0.002000,***
}\hTwoInteract

\begin{tikzpicture}
\begin{groupplot}[
  group style={group size=2 by 1, horizontal sep=0.2cm},
  width=0.58\columnwidth,
  height=4.2cm,
  ybar,
  ymode=log,
  log origin=infty,   
  y dir=normal,       
  ymin=0.00001,
  ymax=1.5,
  enlarge x limits=0.2,
  symbolic x coords={Human,GPT 5.2,Gemma,Llama},
  xtick=data,
  x tick label style={rotate=35, anchor=east},
  grid=major,
  axis on top=false,
  ylabel={Partial $\eta_p^2$},
  label style={font=\sffamily\small},
  tick label style={font=\sffamily\small},
  title style={font=\sffamily\small},
]

\nextgroupplot[
  title={H2: Main effect},
]
\addplot table[x=agent,y=eta2]{\hTwoMain};

\addplot[
  only marks,
  mark=none,
  nodes near coords,
  point meta=explicit symbolic,
  every node near coord/.style={
    anchor=south,
    yshift=2pt,
    font=\sffamily\small
  }
] table[x=agent,y=eta2,meta=star]{\hTwoMain};

\nextgroupplot[
  title={H2: Exp $\times$ Affil},
  ylabel={},
  yticklabels=\empty,
  ytick style={draw=none}
]
\addplot table[x=agent,y=eta2]{\hTwoInteract};

\addplot[
  only marks,
  mark=none,
  nodes near coords,
  point meta=explicit symbolic,
  every node near coord/.style={
    anchor=south,
    yshift=2pt,
    font=\sffamily\small
  }
] table[x=agent,y=eta2,meta=star]{\hTwoInteract};

\end{groupplot}
\end{tikzpicture}
\caption{
\textbf{H2: Exposure effects across humans and LLM surrogates.}
Bars show partial $\eta_p^2$ from two-way ANOVA models for (left) the main effect of AI credibility exposure and (right) the exposure $\times$ political affiliation interaction; asterisks denote F-test significance (* $p{<}.05$, ** $p{<}.01$, *** $p{<}.001$).
Humans exhibit a small exposure effect ($\eta_p^2 \approx .003$) with modest moderation ($\approx .002$).
GPT~5.2 matches the human-scale exposure magnitude with minimal interaction, Gemma shows weaker exposure but stronger moderation, and Llama exhibits dramatically inflated exposure and moderation.
All models detect exposure (H2), but calibration varies sharply.
}

\label{fig:h2}
\end{figure}
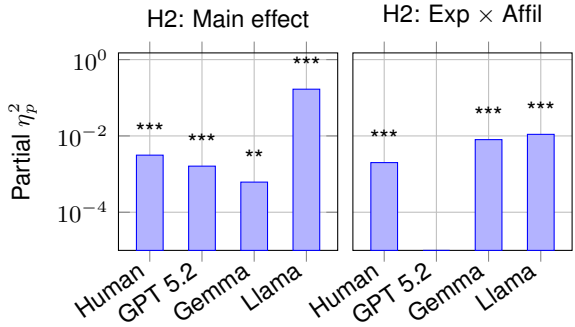

\subsection*{Finding 2: Exposure to AI Credibility Feedback Shifts Ratings (H2)}

We examine whether AI credibility feedback changes perceived headline accuracy and whether this effect varies by political affiliation. Exposure effects are estimated using two-way ANOVA with factors Exposure (Control vs.\ Feedback) and Affiliation (Liberal, Moderate, Conservative). Effect magnitudes are reported as partial $\eta^2$.

\paragraph{Human Results.}

Exposure to AI credibility feedback produces a statistically significant increase in perceived accuracy (Figure~\ref{fig:h2}), with a small but robust effect size ($\eta_p^2 \approx .003$). The exposure effect varies modestly across political groups, yielding a small exposure $\times$ affiliation interaction ($\eta_p^2 \approx .002$). Overall, feedback shifts ratings in humans with limited ideological moderation.

\paragraph{Model Fidelity to Exposure Effects.}
Figure~\ref{fig:h2} compares exposure magnitudes and interaction structure across models. Fidelity differs substantially:

\begin{enumerate}[leftmargin=*, labelwidth=2.6cm, labelsep=0.5em, align=left,
                  itemsep=0pt, parsep=0pt, topsep=2pt]
\item[\textbf{GPT~5.2}] Matches the human-scale exposure effect ($\eta_p^2 \approx .002$) and shows negligible interaction, closely aligning with the human pattern.

\item[\textbf{Gemma}] Exhibits a weaker exposure shift ($\eta_p^2 \approx .001$) but substantially stronger moderation ($\eta_p^2 \approx .008$), indicating amplified ideological structuring.

\item[\textbf{Llama}] Shows dramatically inflated exposure ($\eta_p^2 \approx .168$) and strong moderation ($\eta_p^2 \approx .011$), far exceeding human sensitivity to feedback.
\end{enumerate}

\paragraph{Summary.}
H2 is supported in humans: AI credibility feedback produces a small but reliable increase in perceived accuracy with modest ideological moderation. All surrogate models reproduce the direction of the exposure effect, but magnitudes differ sharply. GPT~5.2 most closely matches the human-scale response, Gemma underestimates overall exposure while amplifying ideological differentiation, and Llama substantially exaggerates feedback sensitivity. These results again reveal pronounced heterogeneity in structural fidelity across models.

\begin{figure*}[t]
\centering
\pgfplotstableread[col sep=comma]{
human,model
2.6268498942917544,0.7869978858350954
1.898113208,0.8084905660377357
1.806924882629108,0.829225352
-1.065274841,-0.610993658
-1.995283019,-0.776100629
-1.151995305,-0.583626761
2.1493128964059203,0.2515856236786469
1.4287735849056604,0.34937106918238925
2.791666666666666,0.5278755868544609
1.6937103594080334,0.4640591966173364
0.9017295597484276,0.4268867924528301
1.968309859,0.5187793427230041
2.857029598308668,0.5930232558139537
2.487264150943396,1.2316037735849052
3.759389671361503,1.070422535211268
-0.344344609,-1.256871036
-1.346383648,-0.345283019
-1.734448357,-0.568661972
-3.16807611,-1.339059197
-3.591509434,-0.912264151
-4.737382629,-0.65258216
-2.334830867,-1.835623679
-3.509433962,-1.472955975
-2.905516432,-1.103579812
-0.429968288,0.014270613
-0.15172956,0.2968553459119496
-0.836267606,0.25880281690140805
1.9828224101479917,0.5438689217758981
1.7768867924528298,0.9746855345911944
2.403462441314553,0.767605634
3.5290697674418605,0.5359408033826636
2.585062893081761,1.4029874213836475
3.265551643192488,1.104753521126761
-3.00845666,-1.628964059
-3.365880503,-1.43490566
-3.638204225,-1.364143192
-1.790433404,-1.64640592
-1.463993711,-1.221855346
-1.66314554,-1.145246479
-1.496035941,-1.259778013
-2.852201258,-0.691352201
-2.877934272,-0.151995305
1.053118393234672,0.3522727272727275
0.5625786163522015,1.1268867924528303
0.6335093896713611,1.002053990610329
1.1662262156448202,0.4223044397463003
0.43144654088050327,0.8930817610062896
2.363262910798123,0.853286385
0.44529598308668117,0.071617336
0.23632075471698055,0.5880503144654083
0.519953052,0.3788145539906109
2.028012684989429,0.26215644820296014
2.142295597484276,0.6199685534591204
1.521713615023474,0.606220657
2.0956659619450324,0.3242600422832975
1.5957547169811317,0.3388364779874218
2.473884976525821,0.472711268
3.138213530655391,0.4965644820295978
1.4319182389937106,1.1146226415094347
1.904929577464789,1.068955399061033
2.126057082452432,0.37103594080338276
2.507232704402516,0.853459119
2.348004694835681,0.819835681
}\hThreeGemma

\pgfplotstableread[col sep=comma]{
human,model
2.6268498942917544,1.0829809725158563
1.898113208,1.411006289308176
1.806924882629108,1.061326291079812
-1.065274841,-0.177061311
-1.995283019,0.033647799
-1.151995305,-0.148180751
2.1493128964059203,0.7455073995771668
1.4287735849056604,1.060377358490566
2.791666666666666,0.5625
1.6937103594080334,0.8517441860465116
0.9017295597484276,0.8407232704402512
1.968309859,0.35416666666666696
2.857029598308668,0.668340381
2.487264150943396,0.6847484276729556
3.759389671361503,0.2567488262910791
-0.344344609,-0.006078224
-1.346383648,0.091823899
-1.734448357,-0.021420188
-3.16807611,-0.142706131
-3.591509434,0.023742138364779564
-4.737382629,-0.035504695
-2.334830867,0.1577695560253698
-3.509433962,0.20141509433962224
-2.905516432,0.020833333333333037
-0.429968288,0.47753699788583504
-0.15172956,0.820440252
-0.836267606,0.34507042253521103
1.9828224101479917,0.7909619450317118
1.7768867924528298,1.0831761006289309
2.403462441314553,0.45099765258215996
3.5290697674418605,0.5110993657505283
2.585062893081761,0.6509433962264151
3.265551643192488,0.1875
-3.00845666,-0.206659619
-3.365880503,-0.053773585
-3.638204225,-0.177523474
-1.790433404,-0.028541226
-1.463993711,0.078144654
-1.66314554,-0.021420188
-1.496035941,-0.017706131
-2.852201258,0.17311320754716952
-2.877934272,0.006748826
1.053118393234672,0.7928118393234671
0.5625786163522015,1.0665094339622643
0.6335093896713611,0.33333333333333304
1.1662262156448202,0.9059196617336154
0.43144654088050327,0.9816037735849052
2.363262910798123,0.52758216
0.44529598308668117,0.8874207188160677
0.23632075471698055,1.158018867924528
0.519953052,0.727992958
2.028012684989429,1.0288054968287526
2.142295597484276,1.3084905660377357
1.521713615023474,1.014377934272301
2.0956659619450324,1.1894820295983086
1.5957547169811317,1.3594339622641503
2.473884976525821,0.930164319
3.138213530655391,0.9920718816067655
1.4319182389937106,1.2267295597484278
1.904929577464789,0.67165493
2.126057082452432,1.0190274841437628
2.507232704402516,1.1746855345911946
2.348004694835681,0.756161972
}\hThreeLlama

\pgfplotstableread[col sep=comma]{
human,model
2.6268498942917544,3.905655391120507
1.898113208,3.8154088050314474
1.806924882629108,2.886737089201877
-1.065274841,-1.709038055
-1.995283019,-1.674056604
-1.151995305,-1.649647887
2.1493128964059203,1.2077167019027488
1.4287735849056604,1.2320754716981135
2.791666666666666,1.181338028169014
1.6937103594080334,1.1522198731501057
0.9017295597484276,0.9628930817610062
1.968309859,0.8782276995305169
2.857029598308668,1.604122621564482
2.487264150943396,1.5622641509433954
3.759389671361503,1.514084507042254
-0.344344609,-1.761363636
-1.346383648,-0.013207547
-1.734448357,-0.347124413
-3.16807611,-2.199788584
-3.591509434,-1.702515723
-4.737382629,-1.973298122
-2.334830867,-1.621564482
-3.509433962,-1.383018868
-2.905516432,-1.427816901
-0.429968288,0.5689746300211418
-0.15172956,0.3746855345911948
-0.836267606,0.2995892018779349
1.9828224101479917,1.2674418604651159
1.7768867924528298,1.2577044025157225
2.403462441314553,1.057511737089202
3.5290697674418605,3.5026427061310788
2.585062893081761,3.476729559748428
3.265551643192488,3.1786971830985915
-3.00845666,-4.757399577
-3.365880503,-4.822798742
-3.638204225,-4.776408451
-1.790433404,-2.824260042
-1.463993711,-2.987735849
-1.66314554,-2.968896714
-1.496035941,-2.914376321
-2.852201258,-2.839308176
-2.877934272,-3.092723005
1.053118393234672,0.6059725158562372
0.5625786163522015,0.9748427672955975
0.6335093896713611,0.827171362
1.1662262156448202,0.8395877378435515
0.43144654088050327,1.156603773584905
2.363262910798123,0.6684272300469489
0.44529598308668117,0.9746300211416496
0.23632075471698055,1.1410377358490562
0.519953052,1.139084507042254
2.028012684989429,1.089323467230443
2.142295597484276,1.1433962264150948
1.521713615023474,1.202171361502348
2.0956659619450324,0.9051268498942919
1.5957547169811317,0.6509433962264151
2.473884976525821,0.7937206572769959
3.138213530655391,1.2727272727272725
1.4319182389937106,1.238522012578616
1.904929577464789,0.8482981220657271
2.126057082452432,1.2426004228329814
2.507232704402516,1.5984276729559754
2.348004694835681,1.233274647887324
}\hThreeGPT

\def\mGemma{0.33794939090962095}
\def\bGemma{-0.04825958405292185}

\def\mLlama{0.15350103148806446}
\def\bLlama{0.4943317350575476}

\def\mGPT{0.7978039700921331}
\def\bGPT{-0.15088433468283358}

\begin{tikzpicture}
\begin{groupplot}[
  group style={group size=3 by 1, horizontal sep=0.6cm},
  width=0.42\linewidth,
  height=0.35\linewidth,
  xlabel={Human $\Delta$},
  ylabel={Model $\Delta$},
  grid=major,
  xmin=-4.9, xmax=4.9,
  ymin=-4.9, ymax=4.9,
  axis equal image,
  label style={font=\sffamily\small},
  tick label style={font=\sffamily\small},
  title style={font=\sffamily\small},
]

\nextgroupplot[title={Gemma}]
\addplot[densely dashed, blue] coordinates {(-5,-5) (5,5)};
\addplot[
  very thick,
  draw=orange,
] ({x},{\mGemma*x + \bGemma});
\addplot[
  blue,
  only marks,
  mark=*,
  mark size=1.0pt,
  fill=blue!30!white
] table[x=human,y=model] {\hThreeGemma};
\node[anchor=north west, font=\sffamily\tiny] at (rel axis cs:0.02,0.98)
{$r=.851$};
\node[anchor=north west, font=\sffamily\tiny] at (rel axis cs:0.02,0.90)
{RMSE$=1.56$};

\nextgroupplot[title={Llama},   ylabel={},
  yticklabels=\empty,
  ytick style={draw=none}]
\addplot[densely dashed, blue] coordinates {(-5,-5) (5,5)};
\addplot[
  very thick,
  draw=orange,
] ({x},{\mLlama*x + \bLlama});
\addplot[
  blue,
  only marks,
  mark=*,
  mark size=1.0pt,
  fill=blue!30!white
] table[x=human,y=model] {\hThreeLlama};
\node[anchor=north west, font=\sffamily\tiny] at (rel axis cs:0.02,0.98)
{$r=.716$};
\node[anchor=north west, font=\sffamily\tiny] at (rel axis cs:0.02,0.90)
{RMSE$=1.91$};

\nextgroupplot[title={GPT 5.2},   ylabel={},
  yticklabels=\empty,
  ytick style={draw=none}]
\addplot[densely dashed, blue] coordinates {(-5,-5) (5,5)};
\addplot[
  very thick,
  draw=orange,
] ({x},{\mGPT*x + \bGPT});
\addplot[
  blue,
  only marks,
  mark=*,
  mark size=1.0pt,
  fill=blue!30!white
] table[x=human,y=model] {\hThreeGPT};
\node[anchor=north west, font=\sffamily\tiny] at (rel axis cs:0.02,0.98)
{$r=.861$};
\node[anchor=north west, font=\sffamily\tiny] at (rel axis cs:0.02,0.90)
{RMSE$=1.16$};

\end{groupplot}
\end{tikzpicture}
\caption{
\textbf{H3: Headline-aware belief updating correspondence.}
Points compare human $\Delta$ to model $\Delta$ across headline--affiliation--label cells ($n=63$).
The dashed line is $y=x$; orange lines are model fits.
All models capture the directional structure of belief updating (GPT~5.2 $r=.861$, Gemma $.851$, Llama $.716$), but differ in magnitude calibration: GPT~5.2 is closest to human scale, Gemma compresses shifts, and Llama shows attenuated slope with bias.
}
\label{fig:finding3}
\end{figure*}
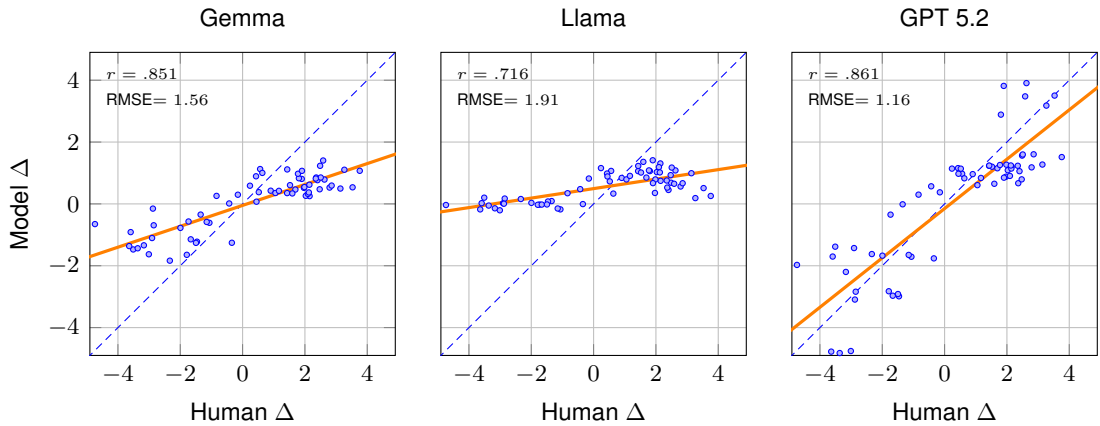

\subsection*{Finding 3: Headline-Aware Belief Updating Structure (H3)}

H3 tests whether belief updating varies systematically across headlines and ideological contexts rather than shifting uniformly.
For each headline–affiliation–feedback cell we compute a headline-aware updating operator
$\Delta =$ Treatment rating minus Control rating.
Figure~\ref{fig:finding3} compares model and human $\Delta$ values.

\paragraph{Human Results.}
Human updating varies substantially across headlines and ideological groups, indicating structured stimulus-level sensitivity.
Belief change depends on both content and alignment context rather than a uniform exposure shift.

\paragraph{Model Fidelity to Belief Updating.}

\begin{enumerate}[leftmargin=*, labelwidth=2.6cm, labelsep=0.5em, align=left,
                  itemsep=0pt, parsep=0pt, topsep=2pt]

\item[\textbf{GPT 5.2}] Closely matches the human updating structure, with points concentrated near the identity line and the strongest correspondence (Figure~\ref{fig:finding3}).

\item[\textbf{Gemma}] Preserves the directional structure but compresses update magnitudes, producing attenuated shifts relative to humans.

\item[\textbf{Llama}] Recovers only coarse structure and exhibits inflated update strength, with greater dispersion from the human pattern.
\end{enumerate}

\paragraph{Summary.}
H3 is supported in human data: belief updating varies systematically across headlines and ideological contexts.
All surrogate models reproduce this structured heterogeneity, but magnitude fidelity differs.
GPT~5.2 most closely matches both the pattern and scale of human updating, Gemma preserves structure with attenuation, and Llama shows reduced alignment with exaggerated shifts.
As in H1 and H2, models capture behavioral structure but differ in sensitivity.

\section{Discussion}

In this paper, we ask whether off-the-shelf LLMs can function as human surrogates in controlled behavioral experiments. Instead of modeling cognition or factual knowledge, we test a simple empirical criterion: if we replace human participants with LLM-generated responses under the same experimental design and statistical analysis, do we reach the same conclusions?

The answer is mixed. Some models reproduce the same directional results and interaction patterns as the human experiment, while others exaggerate or distort effect magnitudes. LLM surrogates can therefore recover whether an effect exists and in which direction, but not necessarily how large it is or how strongly it varies across groups.

To test this question concretely, we used a standard news accuracy rating task in which participants rated the perceived accuracy of political headlines with or without AI credibility feedback.

\subsection{Why AI surrogacy matters}

Human experiments are costly, slow to scale, and logistically constrained by recruitment and participant fatigue. In contrast, LLMs can generate large numbers of responses instantly and at negligible marginal cost. Because these models are trained on extensive human-produced text, their output distributions are often assumed to approximate human judgment distributions at sufficient scale and model fidelity.

This assumption motivates the growing use of LLMs to simulate human responses in survey research, user studies, and policy evaluation. If synthetic responses can substitute for human data, experiments could be conducted faster, at lower cost, and with broader coverage of conditions and populations. However, this substitution is only valid if LLM-generated samples lead to the same empirical inferences as real participants. Otherwise, simulated experiments risk producing misleading conclusions about treatment effects or group differences.

\subsection{Do AI surrogates work?}

Agreement with human findings depends strongly on the model. GPT-5.2 most closely matches human effect sizes and interaction patterns. Gemma preserves the same directional relationships but compresses effect magnitude. Llama detects the same directions but substantially inflates responsiveness to feedback and ideological moderation.

Across models, the main pattern is consistent. All systems detect the presence and direction of exposure and alignment effects. However, the size of these effects and the strength of group differences vary substantially. Synthetic data therefore reproduce which hypotheses are supported more reliably than they reproduce the magnitude of behavioral responses.

This distinction is central for evaluating LLM surrogates. A model that preserves directional conclusions but miscalibrates magnitude may still be useful for hypothesis screening or exploratory design. It is not suitable for estimating realistic behavioral effect sizes. Surrogate evaluation must therefore compare not only statistical significance, but also effect magnitude and interaction strength relative to human benchmarks.

\subsection{Implications for LLM-based behavioral simulation}

The findings support a limited but meaningful role for LLMs as generators of aggregate human-like response patterns. Given only persona descriptors, stimulus text, and condition assignment, models produced rating distributions that reproduced several regularities of human belief updating in this task. This suggests that pretrained models encode statistical associations linking demographic identity, political content, and evaluative judgment that are sufficient to approximate population-level behavioral structure~\cite{argyle2023out,aher2023using}.

This does not necessarily imply that individual human judgments are simple or easily predictable. The present results concern aggregate effects across many observations. LLMs may reproduce distributions while diverging substantially at the level of individual responses, a distinction not evaluated here. The apparent tractability of the task therefore reflects compressibility of group-level patterns rather than faithful modeling of individual cognition.

At the same time, cross-model heterogeneity shows that LLM-generated data are not interchangeable with human data. Systems differ in responsiveness to treatment signals and in the strength of demographic moderation. We therefore conclude that surrogate validity cannot be assumed from directional agreement alone and must be empirically verified against human benchmarks for each hypothesis and model~\cite{messeri2024artificial}.

More broadly, these results suggest a practical workflow for LLM-based simulation. Synthetic samples may be useful for rapid hypothesis screening or exploratory design, where directional effects and relative patterns are of interest~\cite{argyle2023out,park2023generative,li2023camel}. Calibrated human data remain necessary for estimating effect magnitudes and drawing substantive behavioral conclusions. In this sense, LLM surrogates are best understood as exploratory tools rather than replacements for human experiments.

\subsection{Limitations}

Several limitations qualify these conclusions. First, agreement with human aggregate patterns does not imply cognitive equivalence. Models may reproduce effects through learned demographic associations rather than belief formation processes~\cite{abdurahman2024perils,kozlowski2025simulating}. Second, validation is demonstrated in only one experimental paradigm, and generalization to other behavioral domains remains unknown. Third, effect-size comparisons depend on response scaling properties that may differ between human and model distributions despite normalization. Finally, the study evaluates a single prompt encoding of persona and condition; robustness to prompt variation and alternative persona representations remains to be tested.

\section{Conclusions}

We evaluated whether off-the-shelf LLMs can function as human surrogates in controlled behavioral experiments. Given only persona descriptors, experimental condition, and stimulus text, models generated synthetic responses that were analyzed identically to human data. Across systems, LLM surrogates consistently reproduced the direction of key treatment effects but varied substantially in effect magnitude and ideological moderation.

These results indicate that LLMs can approximate aggregate behavioral structure under controlled conditions, but do not reliably recover calibrated behavioral effect sizes. Surrogate validity therefore depends on the model and the hypothesis being tested. Synthetic samples may be suitable for detecting whether an effect exists and for comparing relative patterns, but not for estimating realistic behavioral magnitudes.

LLM surrogates are thus best understood as complementary tools for hypothesis screening and exploratory simulation rather than replacements for human participants. Their appropriate use requires empirical validation against human benchmarks for each experimental domain and model.

\paragraph{Future work}

Future research should evaluate surrogate validity across a wider range of behavioral paradigms, including tasks with richer interaction, longitudinal dynamics, and nonverbal or contextual cues. A key open question is whether LLMs reproduce only aggregate response distributions or also approximate individual-level variation and consistency. Systematic comparisons across prompt formulations, persona representations, and model families will be necessary to determine when synthetic participants provide reliable experimental inference.

\bibliography{ms}

\appendix

\section{Human Participant Data Collection Details}
\label{app:human_data}

This study includes human behavioral data collected through an online survey experiment.

\paragraph{Recruitment.}
Participants were recruited via Prolific. Eligibility criteria restricted participation to individuals located in the United States who self-reported fluency in English. Prolific prescreening filters were used to enforce these criteria prior to enrollment. Participants accessed the experiment through an interactive Web interface.

\paragraph{Compensation.}
Participants received a fixed payment of \$2.00 for approximately 15 minutes of participation, consistent with or exceeding Prolific's recommended minimum hourly compensation rates at the time of data collection. 

\paragraph{Consent Procedure.}
Informed consent was obtained electronically prior to participation. The first page of the survey presented a consent form describing the purpose of the research, the nature of the task, expected duration, compensation, data handling procedures, voluntary participation, and the right to withdraw at any time without penalty. Participants indicated consent by selecting an agreement option before proceeding to the experimental task. Only consenting individuals were allowed to continue.

\paragraph{Ethics Approval.}
The study protocol, materials, and recruitment procedures were reviewed and approved by the authors’ institutional ethics review board prior to data collection.

\paragraph{Data Collected.}
The dataset includes perceived-accuracy ratings of political news headlines under experimentally assigned conditions, along with self-reported demographic attributes (including political affiliation) used for analysis. No personally identifying information was collected.

\end{document}